\documentclass[%
preprint,
 amsmath,amssymb,
 aps,
prc, showpacs,
]{revtex4-2}

\usepackage{graphicx}% Include figure files

\usepackage{dcolumn}% Align table columns on decimal point
\usepackage{bm}% bold math
\usepackage{epsfig}
\usepackage{subcaption}
\usepackage{booktabs}
\begin{document}

%\preprint{APS/123-QED}

\title{Energy dependent ratios of level-density parameters in superheavy nuclei}% Force line breaks with \\
\author{A. Rahmatinejad, T. M. Shneidman, G. Adamian, N. V. Antonenko} \email{antonenk@theor.jinr.ru}
\affiliation{Joint Institute for Nuclear Research, Dubna, 141980, Russia}
\author{P.~Jachimowicz}
 \affiliation{Institute of Physics,
University of Zielona G\'{o}ra, Szafrana 4a, 65516 Zielona
G\'{o}ra, Poland}
\author{M.~Kowal} %\email{michal.kowal@ncbj.gov.pl}
\affiliation{National Centre for Nuclear Research, Pasteura 7, 02-093 Warsaw, Poland}

\begin{abstract}
The nuclear level densities and level-density parameters in fissioning nuclei at their saddle points of fission barriers - $a_{f}$, as well as those for neutron - $a_{n}$, proton - $a_{p}$ , and $\alpha$-particle - $a_{\alpha}$ emission residues at the ground states are calculated for isotopic chains of superheavy nuclei with $Z$=112-120. The calculations are performed with the superfluid formalism using the single-particle energies
obtained from the diagonalization
of the deformed Woods-Saxon potential. Spectra were generated at global minima
of the adiabatic potential energy surfaces,
found by the multidimensional minimization method, and at the proper saddle points, found by the "immersion water flow" technique on multidimensional energy grids, with allowed the reflection and axial symmetry breaking.
The influence of shell effects on the energy dependence of the ratios of level-density parameters corresponding to residues of the considered decay modes to those of neutron emission is studied.
We have shown that, in contrast to the $a_{f}/a_{n}$ ratio, the $a_{p}/a_{n}$ and $a_{\alpha}/a_{n}$ ratios do not show characteristic maxima depending on the excitation energy of the compound nucleus being formed. In the case of alpha decay, we identified the collective enhancement caused by cluster degrees of freedom to play quite an important role. The energetic course of the variability of the level density parameters before reaching the asymptotic value, not taken into account so far, will be of great importance for the estimation of the probabilities of de-excitation cascades via light particles emission in competition with splitting and thus for the determination of the survival probabilities and finally for the total production cross-sections of superheavy nuclei in channels with their (light particles) participation.
\end{abstract}

\maketitle

%\tableofcontents

\section{\label{sec:level1}Introduction}
The experiments on complete fusion reactions with $^{48}$Ca beams and actinide  targets  were successfully carried out at FLNR (Dubna), GSI (Darmstadt), and LBNL (Berkeley)
in order to synthesize superheavy nuclei (SHN) with the charge numbers $Z=112-118$  \cite{Og1,Og2,Og1n,SH1,SH2,SH3,SH4}.
The formation of high-$Z$ evaporation residues in the complete fusion reactions can be roughly treated as a three-stage process: capture, fusion, and survival against fission and charged particle emissions. To analyze the competition between different decay processes in the hot compound nucleus and to calculate evaporation residues cross-sections, one should reliably know the level densities of the final states in various decay modes. This includes the level densities of the nuclei produced after neutron, proton, or $\alpha$-particle emissions and the level densities of the fissioning nucleus at the saddle point (SP) of the fission barrier.

The nuclear level densities (NLD) can be calculated with sophisticated combinatorial models either built on mean-field theory \cite{Hilaire2006,Hilaire2001} or taking into account effective nucleon-nucleon interactions \cite{Alhassid2000,Alhassid2007}. In Ref.~\cite{Goriely2008}, explicitly treating parity, angular momentum, pairing, and shell correlations as well as rotational and vibrational collective enhancements, the combinatorial method has been applied to estimate NLD at the fission SP of actinides.
A recently developed combinatorial method \cite{Ward2017} has been applied for treating the energy and angular-momentum dependencies of nuclear shape evolution in the fission process.
Another technique is the direct microscopic calculation of NLD with the shell model Monte Carlo method used in Ref.~\cite{Alhassid2015} for the iron region and the rare-earth nucleus $^{162}$Dy.
Unlike the mentioned models \cite{Hilaire2006,Hilaire2001,Alhassid2000,Alhassid2007,Goriely2008,Ward2017,Alhassid2015}, in which large-scale computations are required, the thermodynamical models based on the partition function method provide a simpler formalism. With its analytical formula, the Fermi-gas model remains the simplest and widely used model for the calculation of level densities \cite{Bethe1937}. From the Fermi-gas prescription for NLD, the level-density parameter $a$ is a crucial factor indicating the relative importance of each decay mode.
In many applications, the average value of the level-density parameter is often assumed to depend linearly on the mass number $A$ \cite{Sokolov1990}. But in a nucleus that has an internal structure, the level-density parameter depends on the distribution of levels or shell effects. Damping of shell effects with increasing energy causes the level-density parameter to reach its asymptotic value.
In the cases of large shell correction at the ground state (GS), the shell effect in $a$ persists up to high energies.
Within the thermodynamic superfluid model \cite{Rahmatinejad2021}, the level densities of superheavy nuclei (SHN) at the ground state and the SP were systematically studied and the energy and shell-correction dependencies of level-density parameters were compared with those of phenomenological approach \cite{Ignatyuk1975}.
Using all binding energies, fission thresholds as well as shell corrections at  the GS and  SP  from the microscopic-macroscopic model \cite{Jach2021},
it has been shown in Ref.~\cite{Rahmatinejad2021} that the ratio of the level-density parameter at the  SP  to the one at the  GS  is crucial for the survival probability of excited SHN.
The present work is mainly dedicated to the analysis of energy dependence of the relative level-density parameters corresponding to the residues of different decay channels including neutron, proton, and $\alpha$-particle emissions, at their GS as well as fission mode at the SP with using the same mass tables \cite{Jach2021}. Isospin and shell-correction dependencies of level-density parameter ratios at given excitation energies are compared in different decay channels.
For this aim, the level densities and level-density parameters for SHN with $Z$=112-120 are calculated within the thermodynamic superfluid formalism \cite{Decowski1968,origin1} which allows us to take into account the temperature dependence of pairing and shell effects. Application of this formalism was examined in Ref.~\cite{Rahmatinejad2020} through the comparison with the available experimental level densities for Dy and Mo isotopes. Previously the superfluid model was applied to study the level densities of SHN and dinuclear systems using the single-particle spectra obtained with the modified two-center shell model \cite{Bezbakh2014,Bezbakh2016}.

\section{\label{sec:level1}Level-density parameter}

In this work, the single-particle energies are obtained in the microscopic-macroscopic model based on the deformed single-particle Woods-Saxon potential \cite{Cwiok1987} and Yukawa-plus exponential macroscopic energy \cite{Krappe1979} with parameters specified in \cite{Muntian2001}.
Both non-axial and mass-asymmetrical shapes are allowed in the parametrization of the shape of the nuclear surface.
SP are found via immersion water flow method on multi-dimensional spaces of deformations while GS by conjugate gradient minimization method.
For odd and odd-odd nuclei, adiabatic potential energy surfaces were calculated by additional minimization over configurations with one blocked neutron or/and proton on a level from the 10th below to the 10th above the Fermi level, for details see \cite{Kowal2010, Jach2014, Jachimowicz2017_I, Jach2021}.
Thus obtained decay thresholds, i.e. fission barriers, separation energies, shell corrections, with an additional estimated Coulomb barriers for charged particles, are taken into account in the next step in which we use the statistical formalism allowing us to estimate the level-density parameters with the deformed single-particle spectra.
Based on the superfluid formalism \cite{origin1} and with an assumption of thermal equilibrium between neutron (the number of neutrons is $N$) and proton subsystems, the constants of the pairing interaction for neutrons ($G_{N}$) and protons ($G_{Z}$) are adjusted to obtain the GS pairing gaps ($\Delta_{N}$ and $\Delta_{Z}$) at zero temperature with Bardeen-Cooper-Schrieffer (BCS) equations:
\begin{equation} \label{eq1}
Z = \sum_k \left(1-\frac{\varepsilon^{Z}_{k}-\lambda_{Z}}{E^{Z}_{k}}\tanh\frac{\beta E^{Z}_{k}}{2}\right),
\end{equation}
\begin{equation} \label{eq1b}
N = \sum_k \left(1-\frac{\varepsilon^{N}_{k}-\lambda_{N}}{E^{N}_{k}}\tanh\frac{\beta E^{N}_{k}}{2}\right),
\end{equation}
\begin{equation} \label{eq2}
\frac{2}{G_{N,Z}}=\sum_k\frac{1}{E^{N,Z}_{k}}\tanh\frac{\beta E^{N,Z}_{k}}{2},
\end{equation}
where $Z$ and $N$ are the numbers of protons and neutrons in the nucleus, respectively.
The quasiparticle energies $E^{N,Z}_{k}=\sqrt{(\varepsilon^{N,Z}_{k}-\lambda_{N,Z})^2+\Delta_{N,Z}^2}$
are obtained using the single-particle energies ($\varepsilon^{N,Z}_{k}$) calculated with the deformed Woods-Saxon potential.
In our calculations, in the cases of $\Delta_{N,Z}<0.25$ MeV the pairing gap is taken as 0.25 MeV.
Using the values of pairing constants obtained, the pairing gaps and chemical potentials $\lambda$ are determined by solving Eqs.~(\ref{eq1}), (\ref{eq1b}), and (\ref{eq2}) at given temperatures $T=1/\beta$.
Then, using these values, the excitation energies $U=U_{Z}+U_{N}$, entropies $S=S_{Z}+S_{N}$ and intrinsic level densities $\rho$ are calculated as
\begin{equation} \label{eq3}
E_{Z,N}(T)=\sum_k\varepsilon_k\left(1-\frac{\varepsilon^{Z,N}_{k}-\lambda_{Z,N}}{E^{Z,N}_{k}}\tanh\frac{\beta E^{Z,N}_{k}}{2}\right)-\frac{\Delta^{2}_{Z,N}}{G_{Z,N}},
\end{equation}
\begin{equation} \label{eq4}
U_{Z,N}(T)=E_{Z,N}(T)-E_{Z,N}(0),
\end{equation}
\begin{equation} \label{eq5}
S_{Z,N}(T)=\sum_{k}\left\{\ln[1+\exp(-\beta E^{Z,N}_{k})]+\frac{\beta E^{Z,N}_{k}}{1+\exp(\beta E^{Z,N}_{k})}\right\},
\end{equation}
\begin{equation} \label{eq6}
\rho=\frac{\exp{(S)}}{(2\pi)^{\frac{3}{2}}\sqrt{D}},
\end{equation}
where
 \begin{equation} \label{eq6b}
D=
\begin{vmatrix}
\frac{\partial^{2}S}{\partial \beta^{2}} & \frac{\partial^{2}S}{\partial \beta \partial \mu_{Z}} & \frac{\partial^{2}S}{\partial \beta \partial \mu_{N}} \\
\frac{\partial^{2}S}{\partial \beta \partial \mu_{Z}} & \frac{\partial^{2}S}{\partial \mu_{Z}^{2}} & 0 \\
\frac{\partial^{2}S}{\partial \beta \partial \mu_{N}} & 0 & \frac{\partial^{2}S}{\partial \mu_{N}^{2}}
\end{vmatrix}
\end{equation}
is the determinant of the matrix comprised of the second derivatives of the entropy with respect to $\beta$ and $\mu_{Z,N}=\beta\lambda_{Z,N}$.
From the condition of thermal equilibrium between neutron
and proton subsystems the terms $\partial S^2/\partial \mu_N \partial \mu_Z$ are assumed to be zero.

\begin{figure}[t!!]
\centering
\includegraphics[width=.5\linewidth] {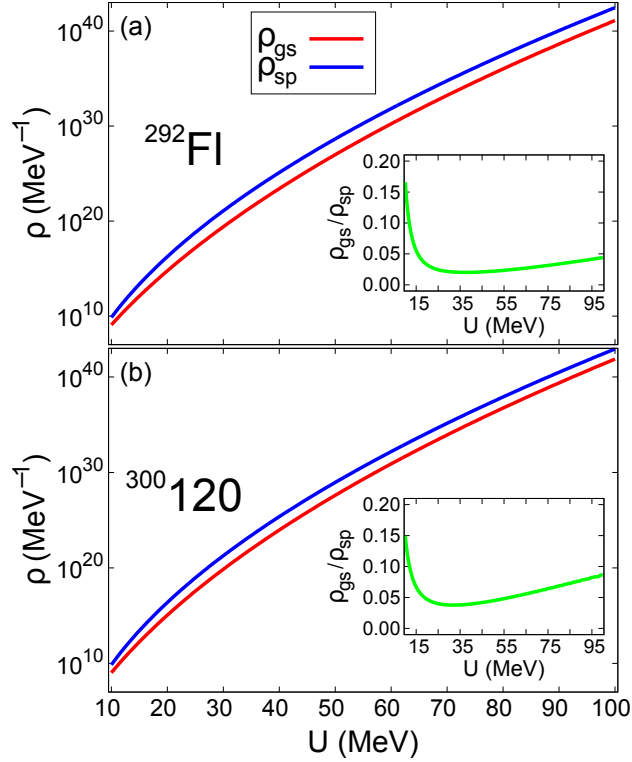}
\caption{Excitation energy dependence of the intrinsic level densities calculated for: (a) $^{292}$Fl, and (b) $^{300}$120
with the single-particle spectra obtained at their GS (red lines) and SP (blue lines).
The inserted panels (green lines) show the ratio of the level densities at the GS to those at the SP.
Here, the excitation energy $U$ is with respect to the corresponding potential energy.}
\label{LDratio2fig}
\end{figure}

The calculations are performed for the isotopic chains of nuclei with $Z$=112-120 to obtain the information needed to estimate the evaporation residues cross sections in neutron, proton, and $\alpha$-particle emission channels. In order to obtain the level densities of nuclei in fission channel, the calculations were performed for the same nuclei using the single-particle energies obtained with the Woods-Saxon potential at the SP. In the BCS calculations at SP, the pairing constants were taken as in the GS.
In Fig.~\ref{LDratio2fig}, the excitation energy dependencies of the intrinsic level densities are presented for $^{292}$Fl, and $^{300}$120, at their fission SP and GS. The ratios of these level densities are shown as well. Because of higher density of single-particle states at the SP, the NLD at the SP are larger than those at the GS at the same excitation energy. Note that at in the calculations of fission width the excitation energies are not the same at the GS and SP. For a comparison between various decays, the energy cost to overcome the fission SP should be taken into consideration.

Fitting the calculated intrinsic level density as a function of excitation energy in the interval between 10 and 100 MeV with the back-shifted Fermi gas expression \cite{Bethe1937}
\begin{equation} \label{eq7}
\rho_{FG}(U)=\frac{\sqrt{\pi}}{12 a^{1/4}(U-\Delta)^{5/4}}\exp({2\sqrt{a(U-\Delta)}}),
\end{equation}
the energy dependent level-density parameter $a(U)$ is obtained. The energy back-shifts in Eq.~\eqref{eq7}
are taken as $\Delta=12/\sqrt{A},0,-12/\sqrt{A}$ for even-even, odd-$A$ and odd-odd isotopes, respectively.
Note that in the cases of particle emissions resulting in the residues with very close values of shell-corrections,
different definition of energy back-shifts can affect the energy dependence of level-density parameter ratios at low energies.
However, it is not the case of the level-density parameter ratios in the energy range of interest $U>25$ MeV.

\begin{figure}
\centering
\includegraphics[width=0.5\textwidth] {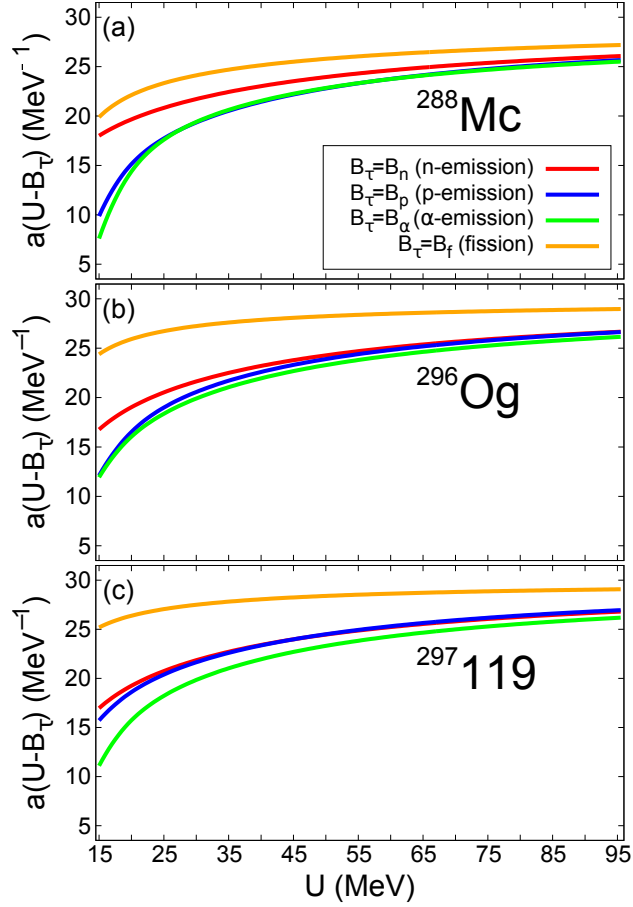}
\caption{Level density parameters obtained for the residual nuclei corresponding to
neutron, proton, and $\alpha$-particle decay channels as well as fission versus excitation energy of mother
nuclei: (a) $^{288}$Mc, (b) $^{296}$Og, and (c) $^{297}119$. $B_f$ is the fission barrier whereas $B_{\tau}$ ($\tau=n,p,\alpha$)
are the particle emission energy thresholds corresponding to each decay channel, respectively.}
\label{aU}
\end{figure}

In Fig.~\ref{aU}, the calculated level-density parameters are shown for the neutron, proton, and $\alpha$-particle decay
channels as well as for fission mode versus the excitation energy of mother nuclei: $^{288}$Mc, $^{296}$Og and $^{297}119$.
The energy costs due to fission barrier $B_{f}$, neutron separation energy $B_{n}=V_{c}^{p}-Q_p$, and  proton $B_{p}$  and $\alpha$-particle $B_{\alpha}=V_{c}^{\alpha}-Q_\alpha$  energy  thresholds, where the Coulomb barriers for charged particles
\begin{eqnarray} \label{eq8}
V_{c}^{p,\alpha}=\frac{1.44(Z-Z_{p,\alpha})Z_{p,\alpha}}{C_{p,\alpha}[(A-A_{p,\alpha})^{1/3}+A_{p,\alpha}^{1/3}]}
\end{eqnarray}
and $Q_{p,\alpha}$-values from the microscopic-macroscopic model \cite{Jach2021}  are taken into consideration.
In Eq.~\eqref{eq8}, $A_p=Z_{p}=1$, $A_{\alpha}=2Z_{\alpha}=4$, and the constants $C_{p,\alpha}$ are taken as $C_{p}=1.7$ fm and $C_{\alpha}=1.57$ fm \cite{Hong2016}.

As pairing effect is expected to be damped out at energies above the neutron separation, the energy dependence of level-density parameters in Fig.~\ref{aU} is mainly explained by the damping of shell correction with excitation energy. The level-density parameters calculated for fissioning nuclei at the SP with smaller shell corrections than in the GS reach their asymptotic values when the energies are already above 30 MeV. However, the shell effects at the GS survive up to energies higher than 50 MeV. As seen in Fig.~\ref{aU}, different damping rates of the shell effects result in particular energy dependent ratios of the level-density parameters in each decay channels. This effect is important to be taken into account in the study of competition between different decay modes at the excitation energies at which the shell effects are not completely damped out.

\section{\label{sec:lev2} Ratios of level-density parameters}

Survival of excited SHN under neutron emission is strongly affected by the ratios of level-density parameters
\begin{eqnarray} \label{eq9}
\frac{a_{f}}{a_{n}}=\frac{a_{sp}(A,U-B_{f})}{a_{gs}(A-1,U-B_{n})},
\end{eqnarray}
\begin{eqnarray} \label{eq10}
\frac{a_{p}}{a_{n}}=\frac{a_{gs}(A-1,U-B_{p})}{a_{gs}(A-1,U-B_{n})},
\end{eqnarray}
and
\begin{eqnarray} \label{eq11}
\frac{a_{\alpha}}{a_{n}}=\frac{a_{gs}(A-4,U-B_{\alpha})}{a_{gs}(A-1,U-B_{n})}.
\end{eqnarray}

\begin{figure}
\centering
\includegraphics[width=0.5\textwidth] {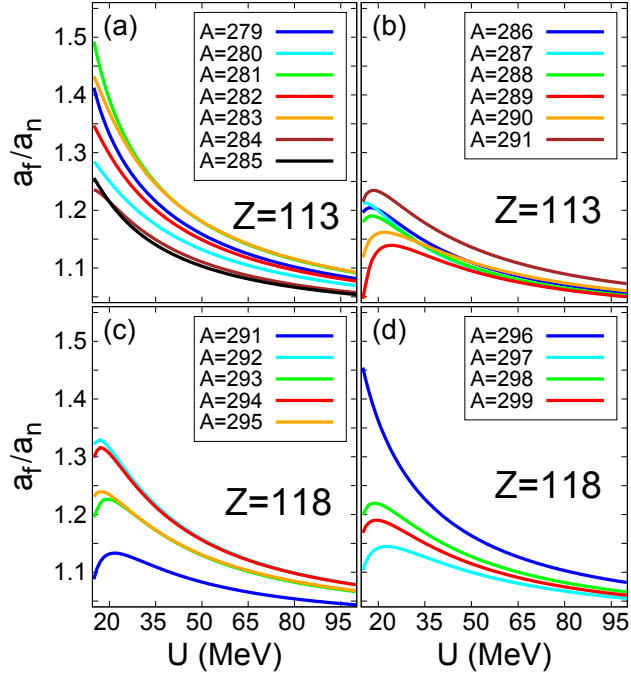}
\caption{Excitation energy dependence of $a_{f}/a_{n}$ calculated from Eq.~\eqref{eq9}
for $^{279-291}$Nh nuclei - panels: (a), (b) and $^{291-299}$Og - panels: (c), (d).
The mass number $A$ of the given isotope is indicated by the selected color.}
\label{aufn}
\end{figure}

The dependencies of $a_{f}/a_{n}$ on excitation energy are displayed in Fig.~\ref{aufn} for $^{279-291}$Nh and $^{291-299}$Og isotopic chains. The value of $a_{f}/a_{n}$ grows fast to a maximum and then slowly falls to asymptotic value less than 1.1. Depending on the difference between the shell-corrections at the SP and at the GS as well as on their damping rates, the values of $a_{f}/a_{n}$ for different isotopes reach maxima at different energies but lower than 25 MeV. The behavior of $a_{f}/a_{n}$ as a function of $U$ depends on the difference between neutron separation energy and the SP height as well as on the shell and pairing effects.
For example, the curve for $^{296}$Og in Fig.~\ref{aufn}~(d) has a maximum at 15 MeV and monotonically decreases with energy to its asymptotic.
As seen in Fig.~\ref{aU}~(b), various energy dependencies of level-density parameters, which is due to different decay thresholds, can lead to a maximum of $a_{f}/a_{n}$ at higher energies. These drawings convince us that the effect of the level density difference: the SP to that of the $A-1$ nucleus will be particularly important for the channel with the emission of one or two neutrons, where the excitation energy of the system is not too high. For example, for the Og, we expect the maximum of excitation function for the $1n$--channel ($2n$--channel) at $U\approx 15$ MeV ($U\approx 25$ MeV).
For these values, the system is far from the asymptotic value where the density parameters converge to a value close to about 1.1.
%Still, it should be remembered that in this energy area one should be careful with the recipe for matching via energy back-shifts formula as Eq.~\eqref{eq7}.

\begin{figure}[t!!]
\centering
\includegraphics[width=0.5\textwidth] {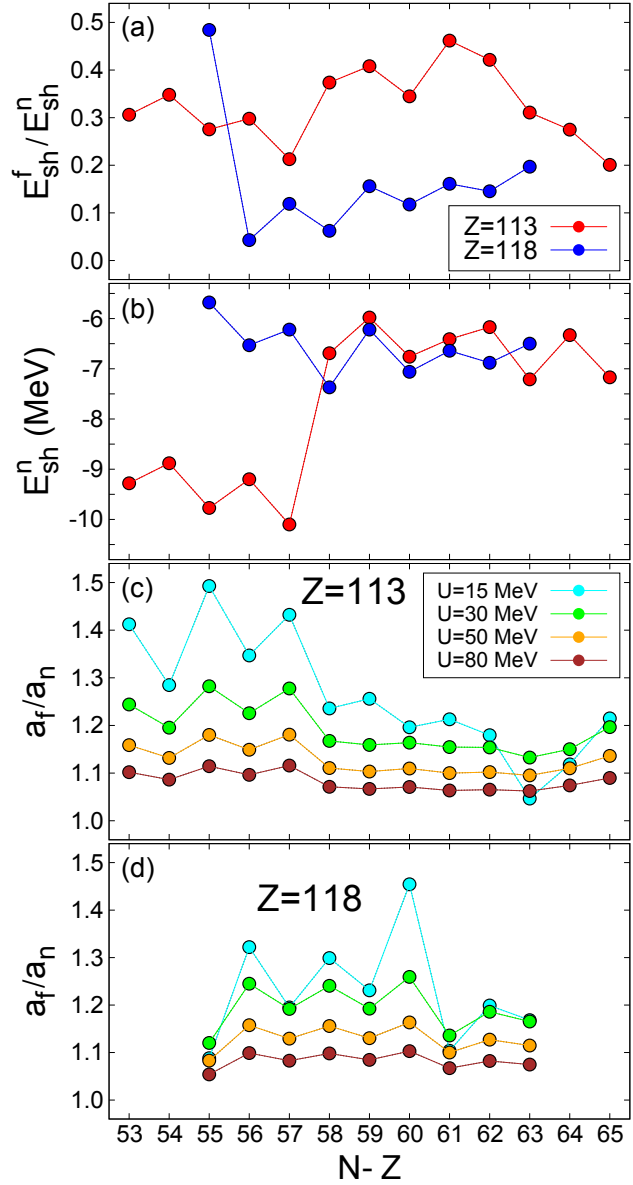}
\caption{Panel (a) - the ratios of shell corrections at the SP $E_{sh}^{f}$
to those at the GS residual nuclei after neutron separation $E_{sh}^{n}$.
Panel (b) - the values of the GS shell corrections of residual nuclei after neutron separation $E_{sh}^{n}$.
The $Z=113$ isotopes are shown by red color while $Z=118$ by blue one.
Panels: (c) and (d) - the level-density parameter ratios $a_{f}/a_{n}$ from Eq.~\eqref{eq9} obtained at 15 MeV (turquoise),
30 MeV (green), 50 MeV (orange), and 80 MeV (brown) for: (c) Nh and (d) Og isotopic chain versus their isospin numbers.}
\label{afn113}
\end{figure}
At $U>30$ MeV the SP shell corrections are mostly damped and the energy dependence of $a_{f}/a_{n}$ is ruled by the GS shell correction. This is seen in Fig.~\ref{afn113} for Nh and Og isotopic chains, respectively. In these figures, the isospin ($N-Z$) dependence of the ratio of shell corrections at the fission SP to those at the GS are displayed in panel (a), the shell corrections at the GS of nuclei after neutron emission are displayed in panel (b), and the ratios $a_{f}/a_{n}$ are shown in panels (c) and (d) versus isospin value.

\begin{figure}
\centering
\includegraphics[width=0.5\textwidth] {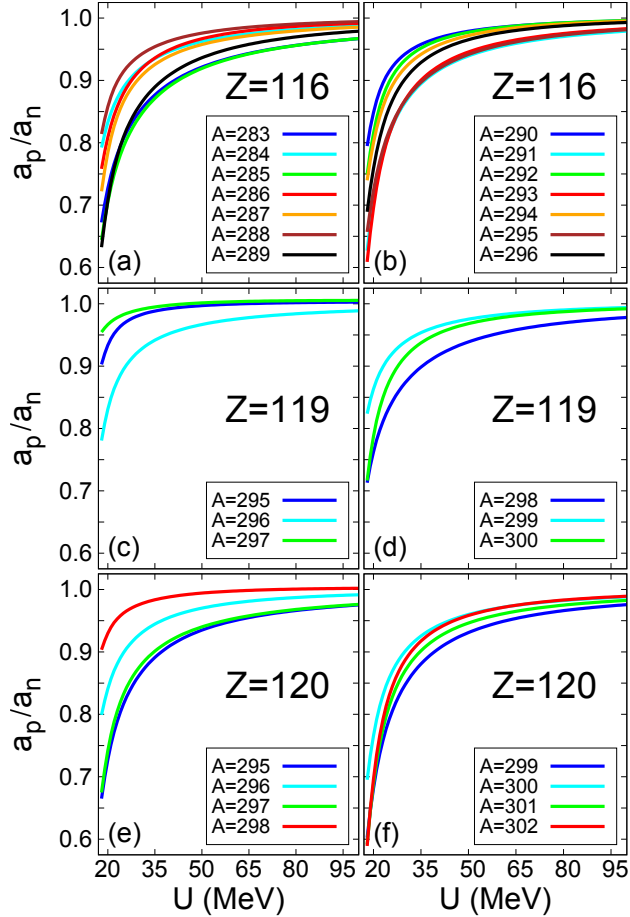}
\caption{Excitation energy dependence of $a_{p}/a_{n}$ calculated with Eq.~\eqref{eq10}
for $^{282-296}$Lv - panels: (a), (b), $^{295-300}$119 - panels: (c), (d), and $^{295-302}$120 - panels: (e), (f).
The mass number $A$ of the given isotope is indicated by the selected color.}
\label{aupn}
\end{figure}

\begin{figure}
\centering
\includegraphics[width=0.5\textwidth] {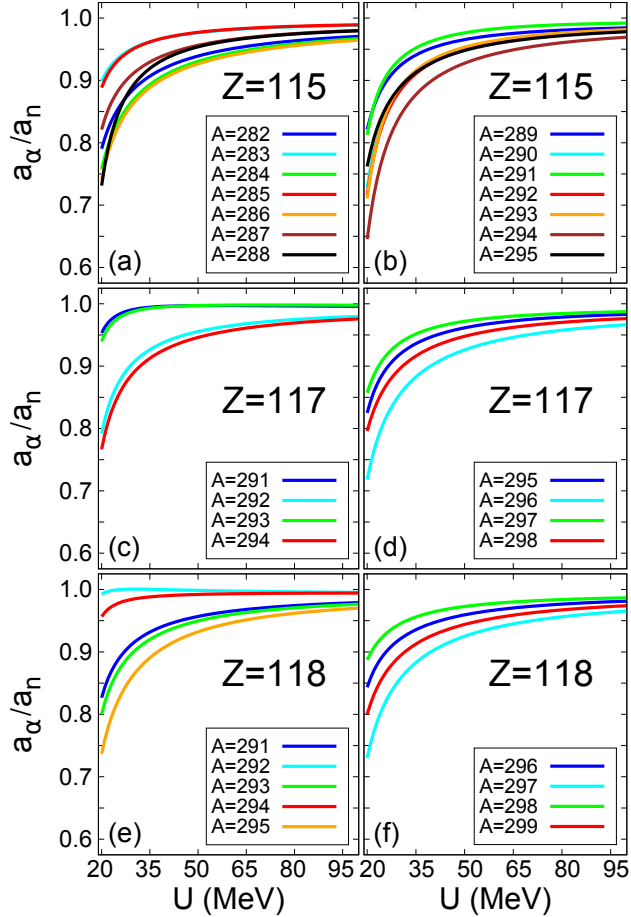}
\caption{Excitation energy dependence of $a_{\alpha}/a_{n}$ calculated with
Eq.~\eqref{eq11} for $^{282-295}$Mc - panels: (a), (b), $^{291-298}$Ts  - panels: (c), (d),
and $^{291-299}$Og - panels: (e), (f). The mass number $A$ of the given isotope is indicated by the selected color.}

\label{aualn}
\end{figure}

In Fig.~\ref{aupn} and \ref{aualn}, the energy dependencies of $a_{p}/a_{n}$ and $a_{\alpha}/a_{n}$ are shown for the isotopic chains of nuclei with $Z=115-120$.
Since the GS have close rates of shell-correction damping, the relative difference between the level-density parameters in particle emission residues uniformly decreases with excitation energy. Therefore, in comparison with $a_{f}/a_{n}$ the energy dependent values of $a_{p}/a_{n}$ and $a_{\alpha}/a_{n}$ do not demonstrate maxima at low energies.
For most of isotopes considered the level-density parameter ratios increase with excitation energy and reach the asymptotic values between 0.95 and 1.

\begin{figure}
\centering
\includegraphics[width=0.5\textwidth] {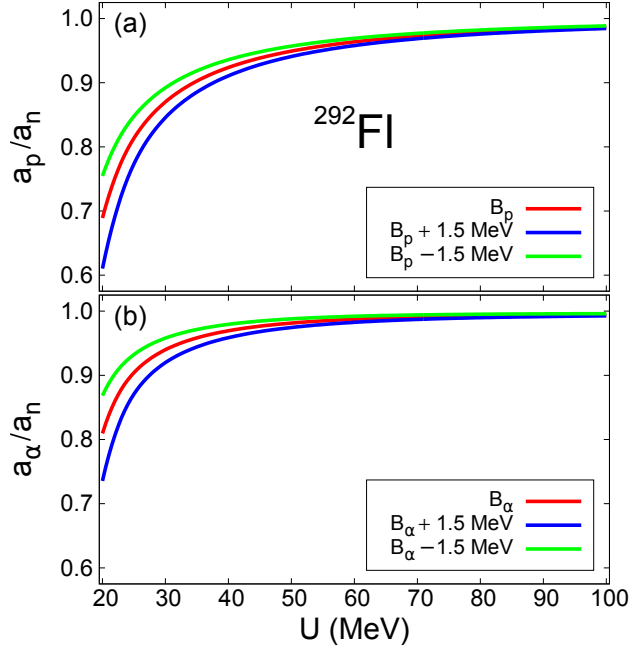}
\caption{Panel (a) - excitation energy dependencies of $a_{p}/a_{n}$ calculated with
Eq.~\eqref{eq10}, and panel (b) - $a_{\alpha}/a_{n}$ calculated with Eq.~\eqref{eq11}, for $^{292}$Fl taking
the particle emission barriers as: $B_{p,\alpha}$ (red lines), $B_{p,\alpha}+1.5$ MeV
(blue lines), and $B_{p,\alpha}-1.5$ MeV (green lines).}
\label{Vdependence2Fig}
\end{figure}

Taking proton emission barrier as $B_{p}\pm1.5$ MeV for $^{292}$Fl, we show the sensitivity of $a_{p}/a_{n}$ to the decay threshold in Fig.~\ref{Vdependence2Fig} (a). As seen, about $\mp 10 \%$ deviation of $a_{p}/a_{n}$ from the one in the case of  proton emission barrier  $B_{p}$ at 20 MeV decreases to zero with increasing excitation energy. The same effect for $a_{\alpha}/a_{n}$ is displayed in Fig.~\ref{Vdependence2Fig} (b). Taking $\alpha$-particle emission barrier as $B_{\alpha}$, and $B_{\alpha}\pm1.5$~MeV, we find about $\mp 8 \%$ difference of $a_{\alpha}/a_{n}$ at $U=20$~MeV that decreases to zero with increasing excitation energy.
At enough high excitation energies the relative importance of charged particles emission with respect to neutron emission is less sensitive to the value of the Coulomb barrier. However, at low excitation energy, especially in multi-step processes in which the energy is reduced by emitted particles, the variation of the Coulomb barrier can play an important role in the estimation of survival probability.

\begin{figure}
\centering
\includegraphics[width=0.5\textwidth] {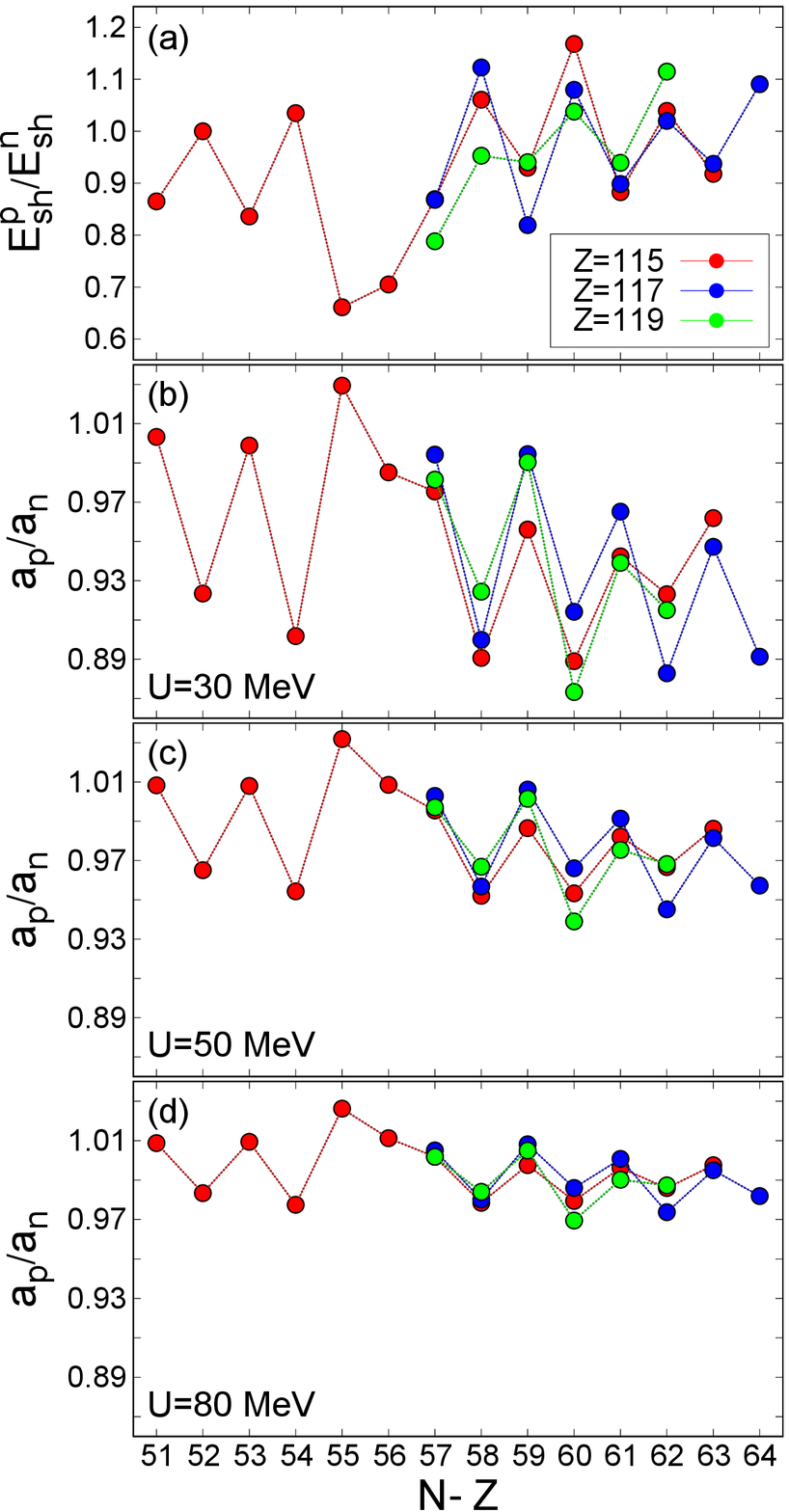}
\caption{For $Z=115$ (red lines), $Z=117$ (blue lines), and $Z=119$ (green lines) isotopic chains, (a) the ratios of shell
corrections corresponding to the residual nuclei after proton emission to those after neutron emission,
and their level-density parameter ratios at: (b) 30 MeV, (c) 50 MeV, and (d) 80 MeV are shown versus their isospin.}
\label{apR-odd}
\end{figure}

\begin{figure}
\centering
\includegraphics[width=0.5\textwidth] {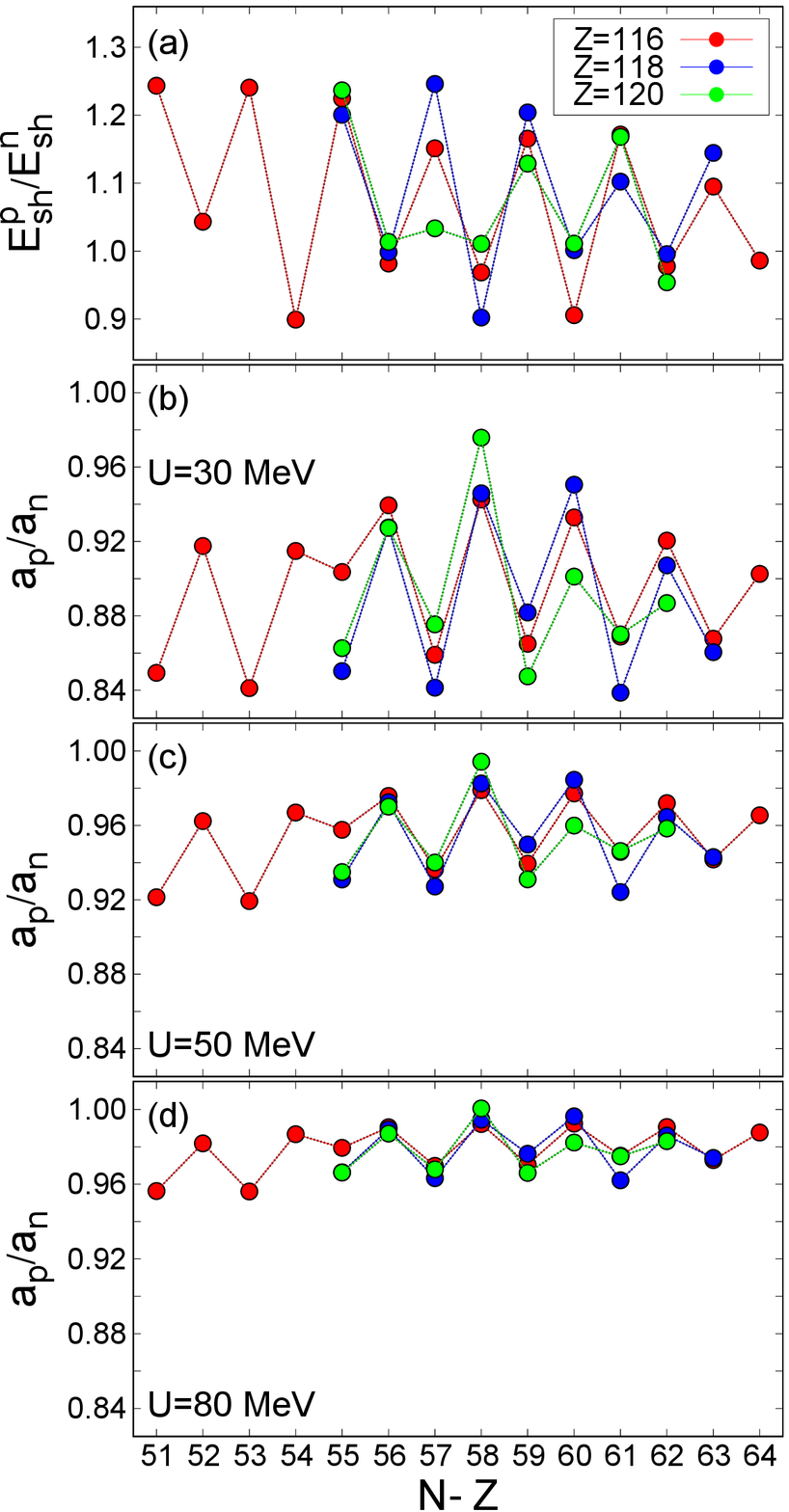}
\caption{The same as Fig.~\ref{apR-odd}, but for $Z=116$ (red lines), $Z=118$ (blue lines), and $Z=120$ (green lines) isotopic chains.}
\label{apR-even2}
\end{figure}

\begin{figure}
\centering
\includegraphics[width=0.5\textwidth] {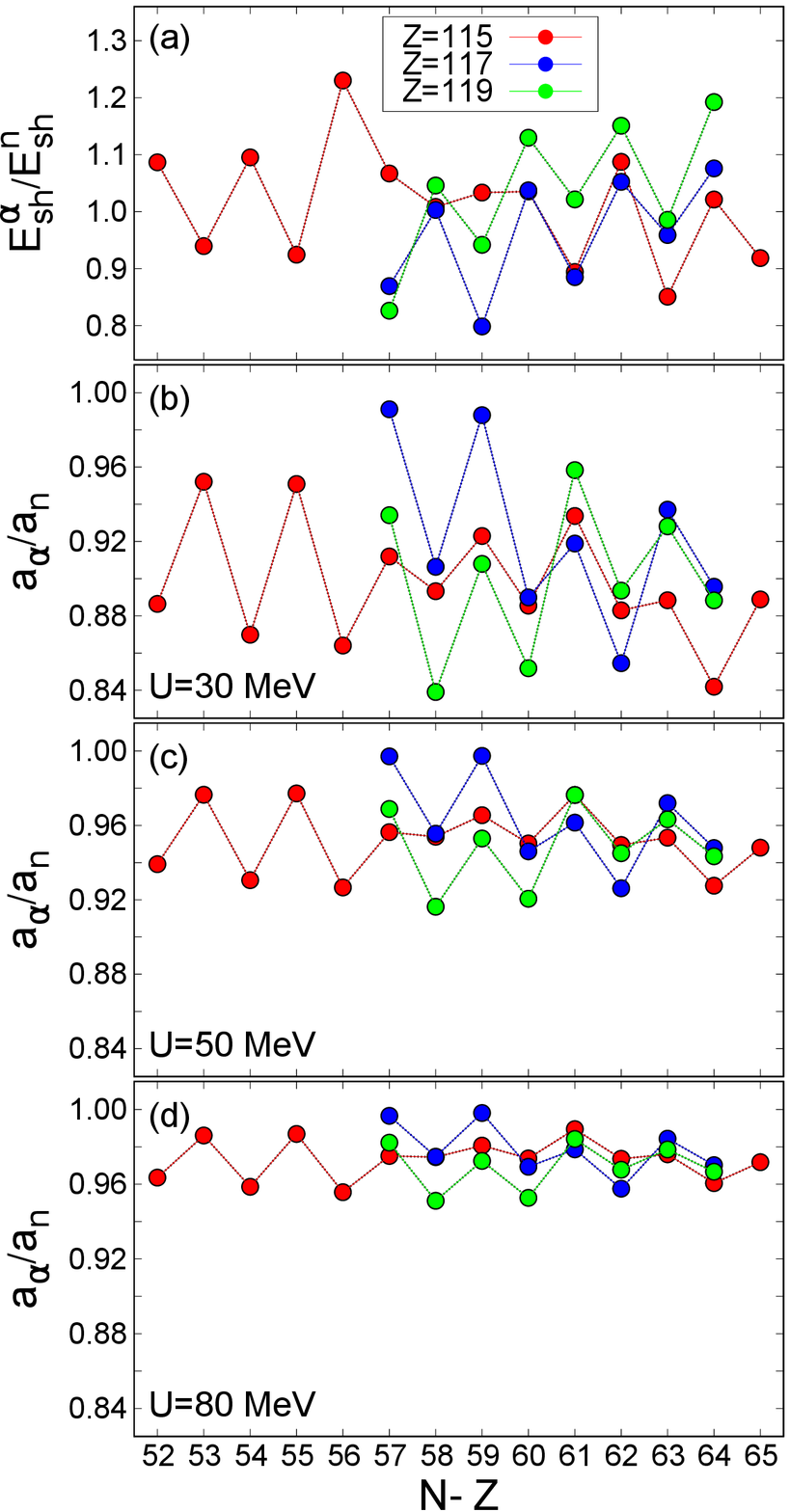}
\caption{For $Z=115$ (red lines), $Z=117$ (blue lines), and $Z=119$ (green lines) isotopic chains,
(a) the ratios of shell corrections corresponding to the residual nuclei after $\alpha$-particle emission to those after neutron emission,
and their level-density parameter ratios at: (b) 30 MeV, (c) 50 MeV, and (d) 80 MeV are shown versus their isospin.}
\label{aalR-odd2}
\end{figure}

\begin{figure}
\centering
\includegraphics[width=0.5\textwidth] {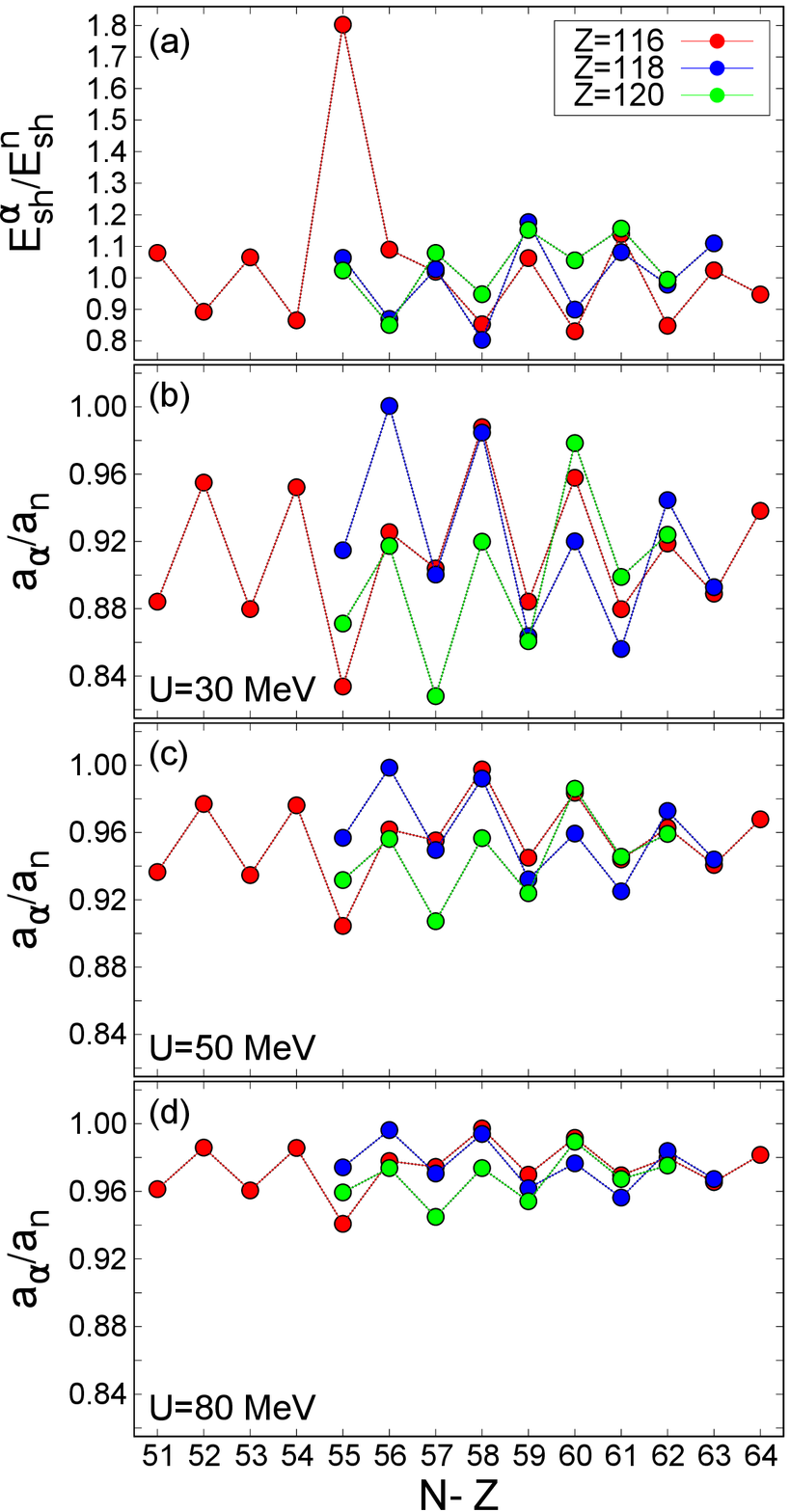}
\caption{The same as Fig.~\ref{aalR-odd2}, but for $Z=116$ (red lines), $Z=118$ (blue lines), and $Z=120$ (green lines) isotopic chains.}
\label{aalR-even2}
\end{figure}

The energy dependence of the level-density parameter ratio is affected by the difference of the decay thresholds and its isospin dependence is mainly affected by the difference of shell-corrections. The relation between the shell-correction ratios (panels a) and the level-density parameter ratios $a_{\alpha}/a_{n}$ and $a_{p}/a_{n}$ (panels b,c, and d) calculated for the isotopic chains of nuclei with $Z=115-120$ is shown in Figs.~\ref{apR-odd}--\ref{aalR-even2} versus isospin. The isospin dependencies presented in Figs.~\ref{apR-odd}--\ref{aalR-even2} for $a_{p}/a_{n}$ and $a_{\alpha}/a_{n}$ remains almost unchanged at $U>30$ MeV. As seen, the positions of maxima of $a_{p}/a_{n}$ and $a_{\alpha}/a_{n}$ coincide with those of minima of $E^{p}_{sh}/E^{n}_{sh}$ and $E^{\alpha}_{sh}/E^{n}_{sh}$, respectively.
This gives an idea how to estimate the relative survival probability under proton ($\alpha$-particle) emission in neighboring isotopes.

Assuming that the collective enhancement of the level densities of residual nuclei are close to each other, we neglect these effects in the calculations in Figs.~\ref{aufn}-\ref{aualn}. So, only the ratios of internal level densities were considered. However, in the case of fission and $\alpha$-emission channels there are additional collective states. In fission they are related to collective rotation and vibration and usually taken into account through the enhancement of level density \cite{Ignatyuk1975}. This enhancement can effectively lead to the ratios $a_{f}/a_{n}$ larger than those in Fig.~\ref{aufn}. So, the phenomenological larger $a_{f}/a_{n}$ effectively accounts for the collective enhancement of level density. If the effect of collective enhancement of level density is well known in fission, we suggest to take into account the collective effects on $\alpha$-emission channel in a similar way.

\begin{figure}
\centering
\includegraphics[width=0.5\textwidth] {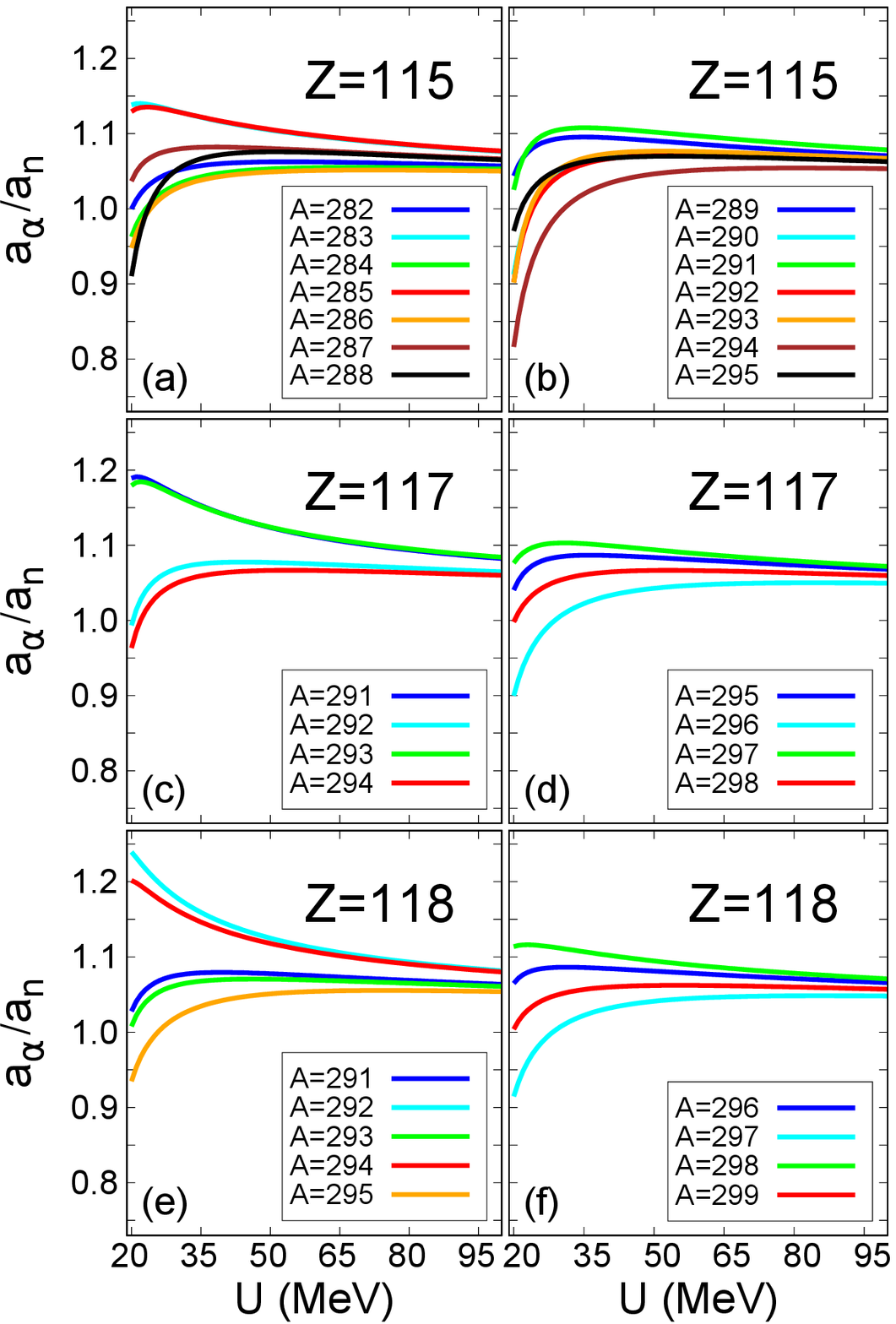}
\caption{Excitation energy dependencies of ratios $a_{\alpha}/a_{n}$ calculated
for $^{282-295}$Mc nuclei - panels: (a), (b), $^{291-298}$Ts - panels: (c), (d), and $^{291-299}$Og - panels: (e), (f),
with account of $K_{\alpha}(\beta)$ from Eq.~\eqref{eq11b} in the level densities of $\alpha$-channel final states.
The mass number $A$ of the given isotope is indicated by the selected color.}

\label{collective}
\end{figure}

The level density for $\alpha$-emission channel can be enhanced because of additional collective states related to the cluster degrees of freedom \cite{Shneidman2015,Shneidman2003}. Assuming the excitations in mass asymmetry motion and relative vibrations of $\alpha$-particle and daughter nucleus, the enhancement factor can be estimated as in Ref.~\cite{Rahmatinejad2020}
\begin{eqnarray} \label{eq11b}
K_{\alpha}(\beta)=\sum_{c}\exp\left(-\beta U_{c}\right)\tau_{c},
\end{eqnarray}
where, $U_{c}=\hbar \omega_{ma} n_{ma}+\hbar \omega_{b}(2n_{b}+|K|)$ and $\tau_{c}=2|K|+1$ are the collective excitation energies and degeneracies, respectively \cite{Shneidman2015,Shneidman2003}. Here, $\hbar \omega_{ma}$ is the frequency of vibrations in mass asymmetry while $\hbar \omega_{b}$ describes the relative vibrations of $\alpha$ particle and daughter nucleus, and $n_{b}$, $n_{ma}$, and $K$ are the corresponding quantum numbers.  Excitation energy dependencies of the ratios $a_{\alpha}/a_{n}$ calculated for $^{282-295}$Mc, $^{291-298}$Ts, and $^{291-299}$Og nuclei with account of $K_{\alpha}(\beta)$ in the level densities are shown in Fig.~\ref{collective}. Taking $\hbar \omega_{ma}=0.5$ MeV, and $\hbar \omega_{b}=0.3$ MeV we find $K_{\alpha}(\beta)$ values varying between 10 and 50, but in the energy range of our interest, one can take effectively into account these collective effects with $K_{\alpha}\approx$20.
This leads not only to an increase in the absolute value of $a_{\alpha}/a_{n}$, but also to a functional change in the dependence of $a_{\alpha}/a_{n}$ on $U$.
For example, the maximum of the ratio $a_{\alpha}/a_{n}$ occurs in the region of energies of 20--30 MeV.
Note that   the ratios $a_{\alpha}/a_{n}$   reach asymptotic values larger than unity.

\section{\label{sec:level14}Summary}

The intrinsic level densities of SHN with $Z=112-120$ at the ground state and at the saddle point are calculated within the thermodynamic  superfluid formalism. The single-particle energies, shell corrections and nuclear masses used in these calculations were obtained within the multidimensional microscopic-macroscopic model \cite{Jach2021}. The energy dependent level-density parameters of the considered nuclei were obtained by fitting the calculated intrinsic level densities by the Fermi-gas expression. Then, the energy dependent ratio of level-density parameters corresponding to the nuclei at the fission saddle point $a_{f}$, and proton $a_p$ and $\alpha$-particle $a_{\alpha}$ emission residues at their ground state to the ones obtained for the daughter nuclei after neutron emission $a_{n}$ were calculated. Generally, the level-density parameter ratios increase with excitation energy and reach an asymptotic value less than 1.1 for $a_{f}/a_{n}$, and less than unity for $a_{p,\alpha}/a_{n}$. The account of collective effects due to cluster degrees of freedom in the level densities of $\alpha$--emission residue enhances the ratio $a_{\alpha}/a_{n}$ to the values larger than unity. So, the values of $a_{\alpha}/a_{n}$ and $a_{f}/a_{n}$ larger than those obtained microscopically effectively account the collective enhancement of level density.

Our analysis shows important effects of decay thresholds and shell correction on the energy dependence of level-density parameter ratios before they reach their asymptotic. Because of large difference in the shell corrections at the saddle point and at the ground state as well as different rates of their damping with excitation energy, the ratios $a_{f}/a_{n}$ have a peak at energy lower than 30 MeV. The saddle-point shell corrections are mostly damped at higher energies and the isospin dependence of $a_{f}/a_{n}$ is mainly related to the ground-state shell corrections of residual nuclei after neutron emission. Channels with the emission of one or two neutrons seem to be particularly sensitive to the change of $a_{f}/a_{n}$, where the available excitation energy does not have to be large and where the system is far from the asymptotic value.
As these channels ($1n$, $2n$) are characteristic of the cold synthesis scenario, it is a strong hint that there also, and perhaps above all, this difference ($a_{f}$ to $a_{n}$) will be very important.

Because of close values of shell corrections and their damping rates at the ground states, the ratios $a_{p}/a_{n}$ and $a_{\alpha}/a_{n}$ monotonically change with excitation energy.

The information and analysis performed in this article can be useful for the calculation of energy dependent ratios of the widths for the considered decay modes and further applied to evaluate corresponding cross sections and probability of survival against fission and charged particle emissions.

\section*{Acknowledgements}
T.M.S, G.G.A., and N.V.A. were supported by the Ministry of Science and Higher Education of the Russian
Federation (Contract No. 075-10-2020-117). M.K. was co-financed by the National Science Centre
under Contract No. UMO-2013/08/M/ST2/00257  (LEA COPIGAL).
The Polish-JINR cooperation program is acknowledged.

\end{document}